\documentclass[aps,pra,preprint,groupedaddress]{revtex4}
\usepackage{amssymb,bm}
\usepackage[dvips]{graphicx}

\begin{document}
\bibliographystyle{apsrev}

\title{Nucleon polarization  in the process $e^+e^-\to N\bar{N}$  near threshold.}
\author{A.E.~Bondar }
%\email{A.E.Bondar@inp.nsk.su}
\author{V.F.~Dmitriev }
\author{A.I.~Milstein  }
\author{V.M.~Strakhovenko}
\affiliation{Budker Institute
of Nuclear Physics\\
and Novosibirsk State University,\\ 630090 Novosibirsk, Russia }

\date{\today}

\begin{abstract}
The process  $e^+e^- \rightarrow N\bar N$ is studied nearby a threshold with account for polarizations of all initial and final particles. The nucleon polarization $\bm \zeta^N$  reveals a strong energy dependence due to that of the nucleon electromagnetic form factors $G_E(Q^2)$ and $G_M(Q^2)$ caused by the final-state
interaction of nucleons. It is shown that the modulus of the ratio of these form factors and their relative phase can be determined by measuring $\bm \zeta^N$ along with the differential cross section. The polarization degree is analyzed  using Paris $N\bar N$ optical potential for calculation of the form factors. It turns out that  $|\bm \zeta^N|$ is high enough in a rather wide energy range above the threshold. Being especially high for longitudinally polarized beams,  $|\bm \zeta^N|$ is noticeable even if both  $e^+e^-$ beams are unpolarized.
\end{abstract}


\maketitle

\section{Introduction}

Experiments with polarized antinucleons may substantially add to our knowledge of the nucleon-antinucleon  ($N\bar{N}$) interaction. However, generating of polarized $\bar{N}$ is a complex task. Very recently it was proposed in Ref.\cite{BDKMST08} to produce polarized $\bar{N}$ at $e^+e^-$ colliders in the reaction $e^+e^-\to N\bar{N}$. For longitudinal polarization of $e^+e^-$ beams considered in Ref.\cite{BDKMST08}, helicity conservation in the annihilation along with the fact that $N\bar{ N}$ pair is produced near the threshold  mainly with zero angular momentum provide high polarization degree of nucleon and antinucleon. Estimates done in Ref.\cite{BDKMST08} with parameters of $c\tau$-factory reported in Ref.\cite{Levichev} show that it is possible to probe a spin-dependent part of the  $N\bar{N}$ inelastic scattering cross section using appropriate polarized targets. The spin-dependent part of the proton-antiproton, $p\bar{p}$, cross section is important at calculations of the polarization buildup rate in the $\bar{p}$ beam interacting with a polarized target in a storage ring (see, e.g., Ref. \cite{DmMilStr08} and references therein).

A recent renewal of interest in low-energy  $N\bar{N}$ physics has been stimulated, in particular, by experimental investigation of the proton (antiproton) electric, $G_E(Q^2)$, and  magnetic, $G_M(Q^2)$,
form factors  in the process $e^+e^- \to p\bar p$ \cite{Bardin94,Armstrong93,Aubert06}. Namely, it was found that the ratio $r=|G_E(Q^2)/G_M(Q^2)|$ strongly depends on $Q^2=4E^2$ ($E$ is the energy in the center-of-mass frame) nearby the reaction threshold at $E=M$. The most natural explanation of this phenomenon is final state interaction of the proton and antiproton, since a strong dependence at small $(E-M)/M$ addresses the interaction at  distances much larger than $1/M$. Therefore, it is possible to use some phenomenological $N\bar{N}$ optical potentials for theoretical description of the process $e^+e^-\to N\bar{N}$ as well as of other processes at low energy (see recent reviews \cite{KBMR02,KBR05}). Such potentials have many parameters which are determined by fitting existing experimental data. The imaginary part of these potentials is known worst of all. Therefore, new data on  $N\bar{N}$ annihilation, especially the spin data, may essentially upgrade quality of theoretical predictions.

In the present paper,  the Paris $N\bar{N}$ optical potential suggested in Ref. \cite{paris82} and then upgraded in \cite{paris94,paris99,Lacombe08} is applied to calculate $G_E(Q^2)$ and $G_M(Q^2)$ by the method used in \cite{DmMil07}. We use the latest version of the Paris $N\bar{N}$ optical potential presented in Ref. \cite{Lacombe08}. Having the form factors at disposal, we analyze the hadron polarization $\bm \zeta^N$ arising in the reaction $e^+e^-\to N\bar{N}$ for arbitrarily polarized $e^+e^-$ beams. It is shown that both the ratio $r=|G_E(Q^2)/G_M(Q^2)|$ and the relative phase of the form factors $\chi$ can be extracted from data on $\bm\zeta^N$ and the differential cross section. In a wide energy range above the threshold, the polarization degree is rather high for transversally and especially for longitudinally polarized beams. It turns out that the polarization is not too small even if both $e^+$ and $e^-$ beams are unpolarized.

\section{Form factors and nucleon polarization}
Using a standard definition \cite{LL} of the electromagnetic hadronic current, one easily obtains
the differential cross section of $e^+e^- \to  N\bar{N}$ annihilation in the center-of-mass frame
\begin{eqnarray}\label{cross}
\frac{d\sigma}{d\Omega}= \frac{\alpha^2\beta S}{2Q^2}|G_M(Q^2)|^2\,,
\end{eqnarray}
where $\beta=\sqrt{1-M^2/E^2}$ is the nucleon velocity, $Q^2=4E^2$, and
\begin{eqnarray}\label{tt}
S = 1+(\bm\nu\cdot\bm \zeta^-)(\bm\nu\cdot\bm \zeta^+)+(R^2-1)\left[\frac{1}{2}(1+\bm \zeta^-\cdot\bm \zeta^+)\bm n_\perp^2
-(\bm \zeta^-\cdot\bm n_\perp)(\bm \zeta^+\cdot\bm n_\perp)\right]\,.
\end{eqnarray}
Here $R=M|G_E(Q^2)|/(E|G_M(Q^2)|)\equiv rM/E$, $\bm n=\bm p/p$, $\bm p$ is the nucleon momentum, $\bm n_\perp=\bm n-\bm\nu(\bm\nu\cdot\bm n)$, $\bm\nu$ is the unit vector parallel to the collision axis; $\bm \zeta^-$ and $\bm \zeta^+$ are polarizations of the $e^-$ and $e^+$ beam, respectively. The cross section (\ref{cross}) is independent of the phase $\chi=\arg(G_E(Q^2)/G_M(Q^2))$. Small terms of the order of $(m_e/E)^2$ ($m_e$ is the electron mass) were neglected in obtaining (\ref{cross}). As a result, the quantity $S$ defined by Eq.(\ref{tt}) vanishes along with the cross section  if both $\bm \zeta^+$ and $\bm \zeta^-$ are collinear with $\bm\nu$ and $\bm \zeta^-\cdot\bm \zeta^+=-1$. So the expression(\ref{tt}) for $S$ is valid if $1+(\bm \nu\cdot\bm \zeta^+)(\bm \nu\cdot\bm \zeta^-)\gg (m_e/E)^2$.

If a polarization of only one of the created particles (nucleon or antinucleon) is measured, the polarization vector $\bm \zeta^N$ reads
\begin{eqnarray}\label{zetaN}
\bm \zeta^N &=& \frac{1}{S}\,(\bm\nu\cdot\bm \zeta^-+\bm\nu\cdot\bm \zeta^+)\{\bm n(\bm \nu\cdot\bm n)+(\bm \nu-\bm n(\bm \nu\cdot\bm n))R\cos\chi\}\\&+&\frac{R}{S}\sin\chi\{
[\bm n\times\bm \nu](\bm \nu\cdot\bm n)(1+\bm \zeta^-\cdot\bm \zeta^+)+[\bm n\times\bm \zeta^+_\perp]
(\bm \zeta^-\cdot\bm n_\perp)+[\bm n\times\bm \zeta^-_\perp](\bm \zeta^+\cdot\bm n_\perp)\}\nonumber\,, \end{eqnarray}
where $\bm \zeta_\perp=\bm \zeta-\bm\nu(\bm\nu\cdot\bm \zeta)$. Since $\bm \zeta^N$ depends on the phase  $\chi$, measurement of the nucleon polarization along with the differential cross section allows one to pin down  the complex function $G_E(Q^2)/G_M(Q^2)$, that is to measure both $r(Q^2)$ and $\chi(Q^2)$.

\begin{figure}[h]
\includegraphics[scale=0.98]{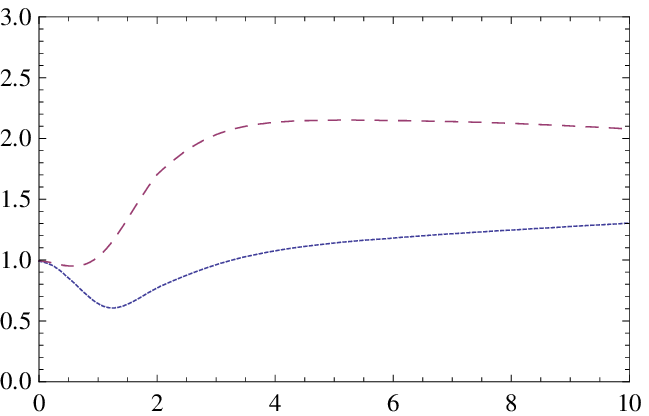}
\includegraphics[scale=0.98]{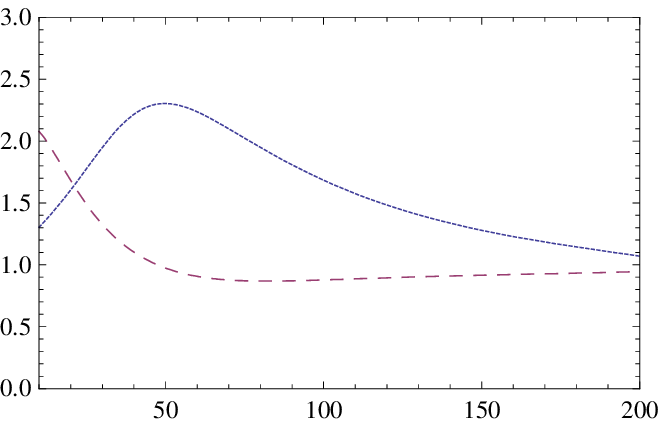}
\begin{picture}(0,0)(0,0)
\put(-200,-10){$T(\mbox{MeV})$}
\put(-390,55){\rotatebox{90}{$r$}}
\end{picture}
\caption{Dependence of  $r=|G_E(Q^2)/G_M(Q^2)|$ on kinetic energy $T=E-M$, $Q^2=4E^2$,  for proton (solid line) and neutron (dashed line).
}\label{r}
\end{figure}

\begin{figure}[h]
\includegraphics[scale=0.98]{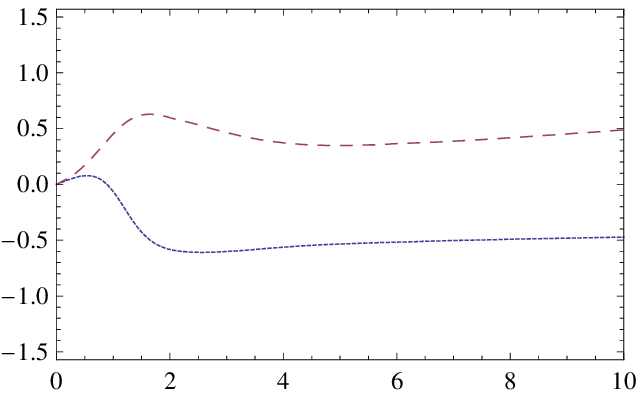}
\includegraphics[scale=0.98]{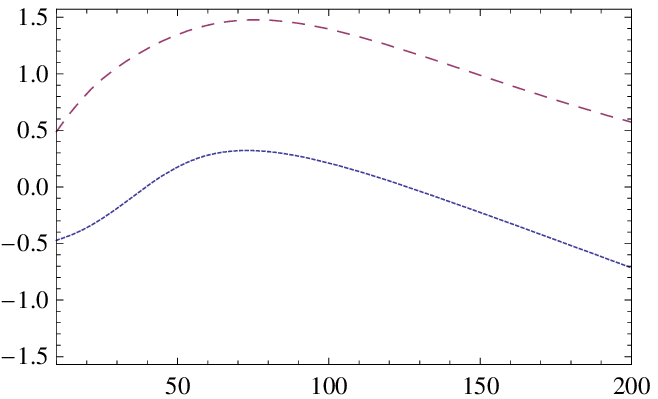}
\begin{picture}(0,0)(0,0)
\put(-200,-10){$T(\mbox{MeV})$}
\put(-380,55){\rotatebox{90}{$\chi$}}
\end{picture}
\caption{Dependence of  the phase $\chi=\arg(G_E(Q^2)/G_M(Q^2))$ on kinetic energy $T=E-M$, $Q^2=4E^2$,
for proton (solid line) and neutron (dashed line).
}\label{chi}
\end{figure}

To find the polarization $\bm \zeta^N$ from Eq.(\ref{zetaN}), one should first estimate the quantities $r(Q^2)$ and $\chi(Q^2)$. Here they are estimated using an approach described in detail in Ref.\cite{DmMil07}, where the Dirac form factors $F_1(Q^2)$ and $F_2(Q^2)$ were considered. The form factors $G_E(Q^2)$ and $G_M(Q^2)$ are expressed via Dirac form factors by (see, e.g., \cite{LL}) relations: $G_E=F_1+\frac{Q^2} {4M^2}F_2$, $G_M=F_1+F_2$. From these relations we have at the threshold $r(Q^2=4M^2)=1$ and $\chi(Q^2=4M^2)=0$. The scheme of calculations is as follows. The bare form factors $\tilde{F}_1$ and  $\tilde{F}_2$ are introduced which address a short-range interaction. They weakly depend on $Q^2$ and
can be considered as phenomenological constants in the vicinity of the threshold. Starting with a seed set of these constants, the dressed form factors ${F}_{1,2}(Q^2)$ are obtained by solving the Schr\"odinger equation for the wave function of the  $N\bar N$ system with the total spin one and isospin one or zero. The $N\bar N$ interaction is described by an optical potential. In that way the cross section of $p\bar p$ and $n\bar n$ production is expressed via bare form factors. The final set of constants $\tilde{F}_1$ and  $\tilde{F}_2$ is that which provides the best fit to experimental data for these cross sections. Using this scheme and the latest version of the Paris optical potential \cite{Lacombe08}, we obtain a dependence of the ratio $r$ and the phase $\chi$ on kinetic energy $T=E-M$  shown in Figs. \ref{r} and \ref{chi} for proton and neutron. In order to stress the strong dependence of these quantities on $T$ in the region close to the threshold, we plot this region on a separate figures.
It turns out that the main uncertainty of our predictions comes from rather big errors in available $n\bar n$ data. Therefore, more accurate measurement of $e^+e^-\rightarrow n\bar n$ cross section would essentially increase the accuracy of theoretical estimates of the form  factors.

The ratio $r$ was also calculated in Ref.\cite{DmMil07} using the previous version \cite{paris99} of the Paris optical potential. Our result for  $r$ is somewhat different from that of Ref.\cite{DmMil07} demonstrating a sensitivity of results to a variation of some parameters of the potential. As is seen in Figs. \ref{r} and \ref{chi}, the quantities  $r$ and $\chi$ reveal a strong dependence on kinetic energy $T=E-M$.

\section{Nucleon  Polarization}

Now we can analyze angular and energy dependence of the polarization $\bm \zeta^N$. From experimental point of view there are three the  most interesting cases: longitudinally  polarized $e^+e^-$ beams, transversally polarized beams, and unpolarized beams.

\subsection{Longitudinally polarized $e^+e^-$ beams}

\begin{figure}[ht]
\includegraphics[scale=0.98]{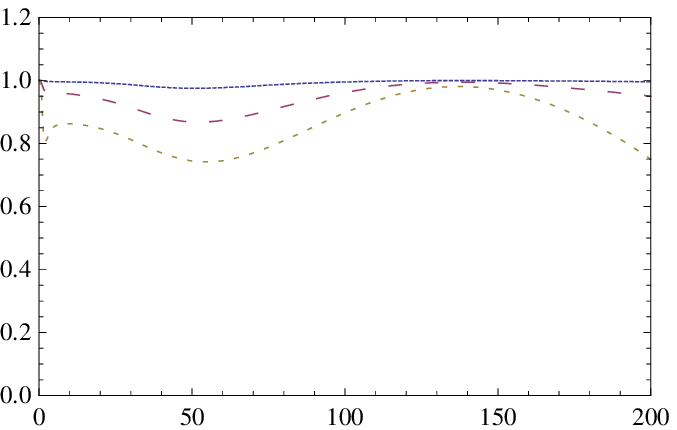}
\hspace{0.5cm}
\includegraphics[scale=0.98]{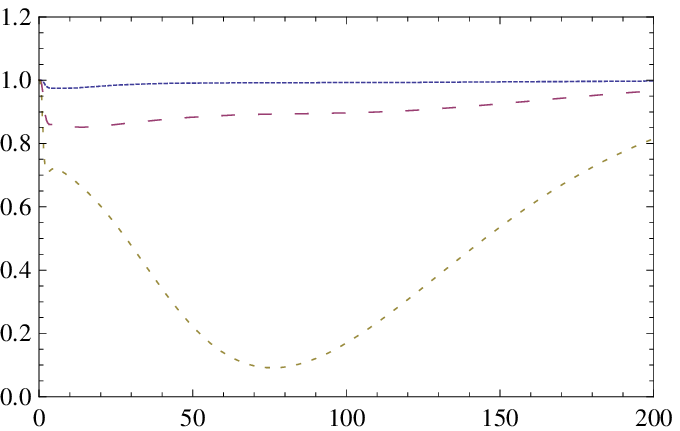}
\begin{picture}(0,0)(0,0)
\put(-130,-10){$T(\mbox{MeV})$}
\put(-340,-10){$T(\mbox{MeV})$}
\put(-430,80){\rotatebox{95}{$\zeta^p$}}
\put(-216,80){\rotatebox{95}{$\zeta^n$}}
\end{picture}
\caption{Dependence of  $\zeta^p$ (left) and  $\zeta^n$ (right ) on kinetic energy $T$ at $b=1$, Eq. (\ref{zetamod}). Angles $\theta$ are: $\pi/8$ (solid line), $\pi/4$ (dashed line), and $\pi/2$ (dotted line).
}\label{zeta}
\end{figure}

For longitudinally polarized beams, where  $\bm\zeta^+_\perp=\bm\zeta^-_\perp=0$, we obtain  from Eqs.(\ref{zetaN}) and (\ref{tt})
\begin{eqnarray}\label{zetaNL}
\bm \zeta^N_l&=&\frac{1}{S_l}\Big\{b[(\bm \nu\cdot\bm n)\bm n+(\bm \nu-(\bm \nu\cdot\bm n)\bm n)R\cos\chi]+
(\bm \nu\cdot\bm n)[\bm n\times\bm \nu]R\sin\chi\Big\}\,,\nonumber\\
S_l&=&1+\frac{1}{2}\sin^2\theta(R^2-1)\, ,\quad
b=\frac{(\bm \nu\cdot\bm \zeta^+)+(\bm \nu\cdot\bm \zeta^-)}{1+(\bm \nu\cdot\bm \zeta^+)(\bm \nu\cdot\bm \zeta^-)}\, ,
\end{eqnarray}
where $\cos\theta=(\bm \nu\cdot\bm n)$. We remind that this formula is valid if $1+(\bm \nu\cdot\bm \zeta^+)(\bm \nu\cdot\bm \zeta^-)\gg (m_e/E)^2$. Since  $N\bar N$ pair is produced near the threshold  mainly with zero angular momentum, the longitudinal polarization of beams and helicity conservation in the annihilation provide high polarization degree $\zeta^N_l=|\bm\zeta^N_l|$ of nucleon and antinucleon. From Eq.(\ref{zetaNL}), we have for $\zeta^N_l$
\begin{eqnarray}\label{zetamod}
\zeta^N_l= \frac{1}{S_l}\left[b^2(\cos^2\theta+ R^2\cos^2\chi\sin^2\theta)+R^2\sin^2\chi\sin^2\theta\cos^2\theta \right]^{1/2}\, .
\end{eqnarray}
This quantity is shown in Fig.\ref{zeta} for $b=1$ at several values of $\theta$.

The polarization degree is rather high in the whole interval of  $T$ considered.

\begin{figure}[ht]
\includegraphics[scale=0.98]{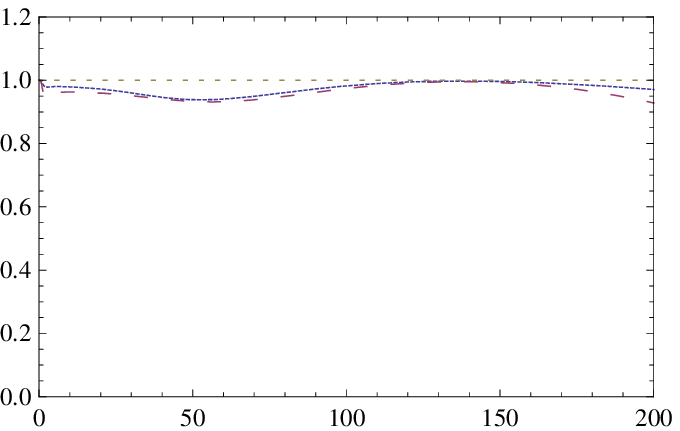}
\hspace{0.5cm}
\includegraphics[scale=0.96]{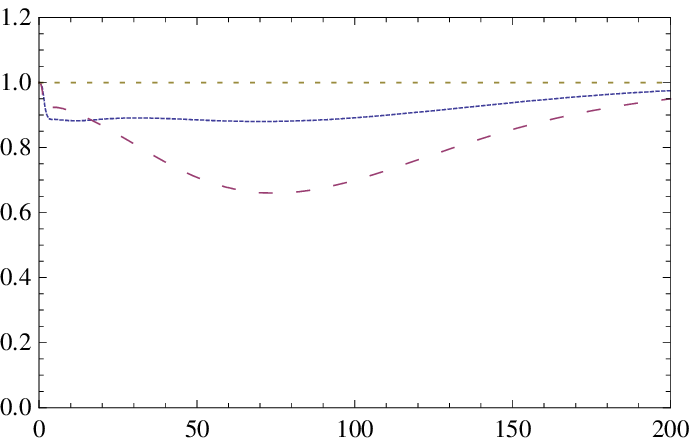}
\begin{picture}(0,0)(0,0)
\put(-125,-10){$T(\mbox{MeV})$}
\put(-340,-10){$T(\mbox{MeV})$}
 \put(-430,60){\rotatebox{95}{$(\bm\zeta_p\cdot\bm\nu)/\zeta_p$ }}
\put(-216,60){\rotatebox{95}{$(\bm\zeta_n\cdot\bm\nu)/\zeta_n$ }}
 \end{picture}
\caption{Dependence of the ratio $(\bm\zeta^p\cdot\bm\nu)/\zeta^p$ (left) and  $(\bm\zeta^n\cdot\bm\nu)/\zeta^n$ (right) on kinetic energy $T$ at $b=1$, Eq. (\ref{nuzetanu}). Angles $\theta$ are: $\pi/8$ (solid line), $\pi/4$ (dashed line), and $\pi/2$ (dotted line).
}\label{zetanu}
\end{figure}

Concerning a direction of $\bm\zeta^N_{l}$, it is collinear with $\bm \nu$ at $\theta=0$ and $\theta=\pi/2$ and does not deviate much from $\bm \nu$ for any $\theta$. This  is seen in Fig.\ref{zetanu} where the quantity $\bm\zeta^N_l\cdot\bm \nu/\zeta^N_l$ is shown. From Eq.(\ref{zetaNL}), we have for $\bm\zeta^N_l\cdot\bm \nu$

\begin{eqnarray}\label{nuzetanu}
\bm\zeta^N_l\cdot\bm \nu =\frac{b}{S_l}[\cos^2\theta+R\cos\chi\sin^2\theta]\,.
\end{eqnarray}
From  Fig. \ref{zetanu}, the direction of polarization $\bm\zeta^N_l$ at $b=1$ is very close to that of $\nu$. So, components of $\bm\zeta^N_{l\perp}$ are small in the whole region of $T$ presented in Fig. \ref{zetanu}.

\subsection{Transversally polarized $e^+e^-$ beams}
\begin{figure}[ht]
\includegraphics[scale=0.98]{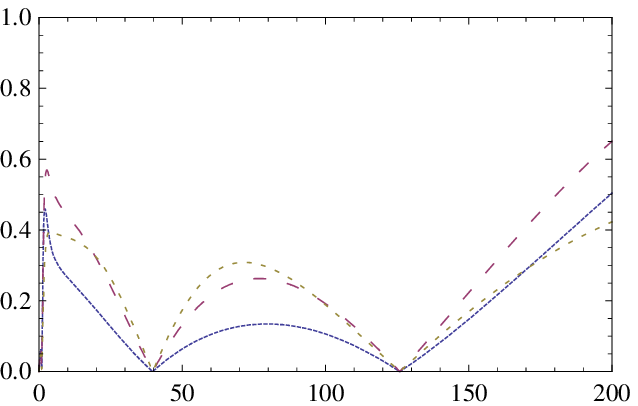}
\hspace{0.7cm}
\includegraphics[scale=0.96]{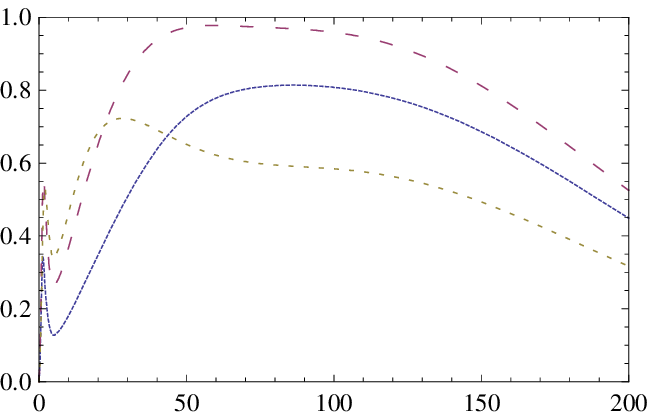}
\begin{picture}(0,0)(0,0)
\put(-125,-10){$T(\mbox{MeV})$}
\put(-340,-10){$T(\mbox{MeV})$}
\put(-420,80){\rotatebox{95}{$\zeta^p_{tr}$ }}
\put(-210,80){\rotatebox{95}{$\zeta^n_{tr}$ }}
\end{picture}
\caption{Polarization degree $\zeta^p_{tr}$ (left) and  $\zeta^n_{tr}$ (right) depending on kinetic energy $T=E-M$, at $\theta=\pi/2$, $a=1$.
Angles $\varphi$ are: $\pi/8$ (solid line), $\pi/4$ (dashed line), and $3\pi/8$ (dotted line).
}\label{zetatran}
\end{figure}

In storage rings $e^+e^-$ beams commonly are polarized due to the radiative polarization. Then beam polarization vectors are transverse with respect to $\bm \nu$. Namely, $\bm \zeta^+=\bm e_z\zeta^+$, $\bm \zeta^-=-\bm e_z\zeta^-$ where the unite vector $\bm e_z$ is directed along a magnetic field and $\bm \nu \cdot \bm e_z =0$. In this case we obtain from Eqs.(\ref{zetaN}) and (\ref{tt})
\begin{eqnarray}\label{zetatr}
\bm \zeta^N_{tr} &=& \frac{R}{S_{tr}}\sin\chi\Big\{
(1-a)[\bm n\times\bm \nu](\bm n\cdot\bm \nu)+2 a[\bm e_z\times\bm n]
(\bm n\cdot\bm e_z)\Big\}\nonumber\,,\nonumber\\
S_{tr} &=&1+\frac{1}{2}(R^2-1)\left[\bm n_\perp^2(1-a)+2 a(\bm e_z\cdot\bm n)^2\right]\,,\quad a=\zeta^+\zeta^-\,.
\end{eqnarray}
Note that $\zeta^N_{tr}$ depends on both angles $\theta$ and $\varphi$, where $\varphi$ is the
angle between $\bm n_\perp$ and $\bm e_z$.  At $\theta=\pi/2$, the vector $\bm \zeta^N_{tr}$ is collinear with $\bm \nu$ for any $\varphi$. Besides, $\bm \zeta^N_{tr}(\theta=\pi/2,-\varphi)=-\bm \zeta^N_{tr}(\theta=\pi/2,\varphi)$ and this property of $\bm \zeta^N_{tr}$  would be helpful in extracting the spin-dependent part of the  $N\bar{N}$ annihilation cross section within a scheme suggested in Ref.\cite{BDKMST08}.
Depending on kinetic energy $T=E-M$, the polarization degree $\zeta^N_{tr}$ is shown in Fig.\ref{zetatran} for proton and neutron at $\theta=\pi/2$, $a=1$, and several values of $\varphi$.
At given $T$ and $\theta=\pi/2$, the polarization degree  $\zeta^N_{tr}$ is a maximum at $\varphi$  defined by the relation $\cos2\varphi=a(1-R^2)/(1+R^2)$. At these points the polarization degree is $$\zeta^N_{tr}=\frac{2R|a\sin\chi|}{\sqrt{(1+R^2)^2-a^2(1-R^2)^2}}$$ Zeros of $\zeta^p_{tr}$ in Fig.\ref{zetatran} are at the points where $\sin\chi$ vanishes (cf. Fig.\ref{chi}).

\subsection{Unpolarized $e^+e^-$ beams}

When both beams are unpolarized ($\bm \zeta^-=\bm \zeta^+=0$), we obtain from Eqs.(\ref{zetaN}) and (\ref{tt})
\begin{eqnarray}\label{zetaunp}
\bm \zeta^N_{un}=\frac{R}{S_l}\sin\chi\cos\theta[\bm n\times\bm \nu]\, ,
\end{eqnarray}
with $S_l$ defined in (\ref{zetaNL}). Note that $\bm \zeta^N_{un}$ is collinear to $[\bm n\times\bm \nu]$ and that the polarization degree $\zeta^N_{un}=|\bm \zeta^N_{un}|$ is independent of the azimuth angle  $\varphi$.
As should be, the expression (\ref{zetaunp}) is reproduced by Eq.(\ref{zetaNL}) at $b=0$ and by Eq.(\ref{zetatr}) at $a=0$. It follows from (\ref{zetaunp}) that, at given kinetic energy $T$, the polarization degree  $\zeta^N_{un}$ is a maximum at $\theta=\theta_0=\arcsin\sqrt{2/(R^2+3)}$ where $\zeta_N^{un}(\theta_0)=R|\sin\chi| /\sqrt{2(R^2+1)}$.
\begin{figure}[ht]
\includegraphics[scale=0.96]{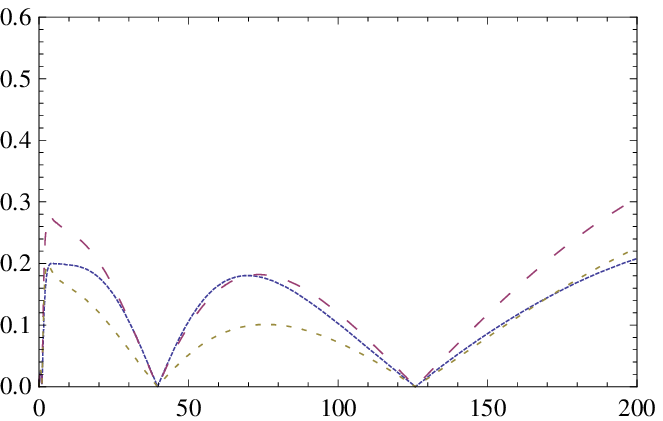}
\hspace{0.7cm}
\includegraphics[scale=0.98]{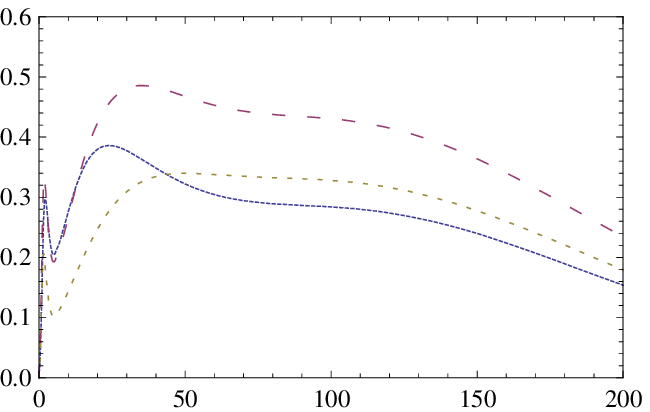}
\begin{picture}(0,0)(0,0)
\put(-125,-10){$T(\mbox{MeV})$}
\put(-340,-10){$T(\mbox{MeV})$}
\put(-415,80){\rotatebox{95}{$\zeta^p_{un}$}}
\put(-210,80){\rotatebox{95}{$\zeta^n_{un}$}}
\end{picture}
\caption{Polarization degree $\zeta^p_{un}$ (left) and  $\zeta^n_{un}$ (right) depending on kinetic energy $T=E-M$, Eq.(\ref{zetaunp}), at angles $\theta$: $\pi/8$ (solid line), $\pi/4$ (dashed line), and $3\pi/8$ (dotted line).
}\label{zetab0}
\end{figure}
Fig.\ref{zetab0} represents a dependence of $\zeta^N_{un}$  on $T$ for proton and neutron at several values of $\theta$. The shape of curves is like  that in the case of transversally polarized beams (cf. Fig.\ref{zetatran}), but now the polarization degree is reduced roughly by half. However, as is seen in Fig.\ref{zetab0},  $\zeta^N_{un}$ is not too small. This fact allows one to measure the phase $\chi$ even if both beams are unpolarized.

\section{Conclusion}

We investigated the energy  dependence of the nucleon polarization $\bm\zeta^N$ in the reactions  $e^+e^-\rightarrow p\bar{p}$ and $e^+e^-\rightarrow n\bar{n}$ near the threshold. For that we calculated the nucleon electromagnetic form factors $G_E(Q^2)$ and $G_M(Q^2)$ using the latest version of the Paris  $N\bar{N}$ potential. In a relatively wide energy range above the threshold the polarization degree is rather high for transversally and especially for longitudinally polarized beams. It turns out that the polarization is not too small even if both $e^+$ and $e^-$ beams are unpolarized. It is shown that the phase $\chi=\arg(G_E(Q^2)/G_M(Q^2))$ and the ratio $r=|G_E(Q^2)/G_M(Q^2)|$ can be extracted from the data on $\bm\zeta^N$ and the differential cross section. In turn, such data would help to diminish uncertainty in the parameters of various potentials which address the $N\bar{N}$ interaction.

We emphasize that formulas obtained above describe production, in  $e^+e^-$ annihilation, of any spin 1/2 baryon-antibaryon pair, as $\Lambda\bar{\Lambda}$ and $\Lambda_c\bar{\Lambda}_c$. Though at present we are unable to calculate $\chi$ and $r$ for $\Lambda\bar{\Lambda}$ system, we expect high polarization degree of baryons in a wide energy region above a threshold if  $e^+e^-$ beams are longitudinally polarized. Let us repeat that the reasons for such expectations are helicity conservation in the annihilation and the fact that $\Lambda\bar{\Lambda}$ pair is produced near the threshold  mainly with zero angular momentum.

Recently, measurements of $\chi$ and $r$ for $\Lambda\bar{\Lambda}$ system were performed in Ref.\cite{Druzh2007} with unpolarized $e^+e^-$ beams  using the initial state radiation technique. Only very weak limits were established in Ref.\cite{Druzh2007} on the relative phase of electromagnetic form factors and the ratio  $r=|G_E(Q^2)/G_M(Q^2)|$ was measured with rather low accuracy due to the limited statistics. Experiment on $c\tau$-factory with longitudinally polarized  $e^+e^-$ beams will provide significantly more accurate information on the baryon electromagnetic form factors in the time-like region. Especially since measurement of $\bm\zeta^\Lambda$ is a more simple experimental task as that for nucleon.

\vspace{1cm}
We are grateful to  I.A. Koop and D.K. Toporkov for fruitful  discussions.
This work was supported in part by the RFBR Grant No 09-02-00024 and the Grant 14.740.11.0082 of Federal Program "Personnel of Innovational Russia".

\end{document}